\documentclass [reprint,pra,aps,showpacs]{revtex4-1}
\usepackage{amssymb, amsthm,amsmath}
\usepackage{graphicx}
\usepackage[dvips]{color}

\def\Var{\mathrm{Var}\,}
\def\Cov{\mathrm{Cov}\,}
\def\<{\langle}
\def\>{\rangle}

\begin{document}

\title{Estimation of covariance matrix of macroscopic quantum states}
\author{L\'aszl\'o Ruppert, Vladyslav C. Usenko and Radim Filip}
\affiliation{Department of Optics, Palacky University, 17. listopadu 12, 771 46 Olomouc, Czech
Republic}
\date{\today}

\pacs{42.50.-p, 03.65.Wj, 42.50.Dv}

\begin{abstract}
For systems analogous to a linear harmonic oscillator, the simplest way to characterize the state is by a covariance matrix containing the symmetrically-ordered moments of operators analogous to position and momentum. We show that using Stokes-like detectors without direct access to either position or momentum, the estimation of the covariance matrix of a macroscopic signal is still possible using interference with a classical noisy and low-intensity reference. Such a detection technique will allow to estimate macroscopic quantum states of electromagnetic radiation without a coherent high-intensity local oscillator. It can be directly applied to estimate the covariance matrix of macroscopically bright squeezed states of light.
\end{abstract}

\maketitle


\section{Introduction} Quantum measurement is the only way to obtain information about quantum systems. In the ideal case, the measurement corresponds to an observable of a quantum system \cite{Peres}.
 A complete set of incompatible observables then allows us to reconstruct the density matrix of the state \cite{Paris}.
 For systems analogous to a linear harmonic oscillator, basic observables are analogous to position and momentum \cite{Sakurai}. 
 A possible characterization of the state based on these observables is by the vector of mean values and the covariance matrix containing symmetrically-ordered second moments of the position and momentum \cite{covmat}. 
The covariance matrix allows us to determine whether the state has squeezed quantum uncertainty lower than that of the ground state in any linear combination of position and momentum  \cite{squeezing1,squeezing2}. 
This is a direct witness of non-classicality of a quantum state (i.e., the state is incompatible with a mixture of coherent states) \cite{noncl}. 
Such non-classical states have been generated in quantum optics and exploited in many quantum information and quantum metrology protocols. 
 Even though the covariance matrix does not completely characterize the state in the general case (unlike the density matrix), it provides sufficient information to completely ensure the security of quantum key distribution with continuous variables \cite{Nav,Gar,Mad12,Rup14} and the high sensitivity of quantum metrology \cite{Morgan1,Luis1,Morgan2,Genovese,Brask1}.

In quantum optics, homodyne detection is a way to directly access in-phase and out-of-phase quadratures of fluctuating electric field \cite{Yuen,Schu}, equivalent to position and momentum observables of light.
This homodyne measurement mixes the measured light with a local oscillator used as a reference.
The local oscillator is a low-noise classical coherent beam with a large intensity, much larger than the measured optical signal. 
This coherently amplifies the measured quantum light, makes linear detection possible and also provides a high-quality phase reference for a phase-sensitive homodyne detector. 
However, recent optical experiments with macroscopically bright squeezed vacuum \cite{Isk1,Isk2,Isk3,Isk4} led to a new interesting situation. 
These non-classical states are so bright that it is possible to directly measure them by ordinary linear intensity detectors for macroscopic light without the necessity of any amplification effect provided by the local oscillator. 
 For the standard homodyne measurement it would be necessary to have a local oscillator as a reference that is even much brighter than the signal \cite{Use}, which could be impractical. 

Instead, to measure these bright states, Stokes-like measurements generalizing the homodyne detection including any state as reference are considered \cite{Stok1,Stok2,Stok3}. In this situation the linearization and the direct detection of quadratures does not work anymore, tomographic methods should be used instead to obtain the signal state from the measurement data provided by a non-linear detector. 
Note that there is a well-known similar scenario in which both the signal and the reference are weak \cite{Vogel,Vogel2,Opat,Sanchez,Jiang}. 
The difference is, however, crucial: in that case one can use photon counter detectors, which provide the exact photon number distribution. But the photon number resolving detectors are not suitable for macroscopic states of light, hence in our case with intensity detectors only the moments of the photon numbers are accessible, which means that the complete density matrix in Fock state representation cannot be estimated.

The Stokes operators correspond to the sum and difference of photon numbers between the signal and the reference modes coupled on a beam splitter. They have been extensively used to measure and quantify the degree of polarization, polarization squeezing and entanglement \cite{pol1,pol2,pol3,pol4,pol5,pol6,Rivas2,pol7,pol8,pol9,pol10,pol11,pol12,pol13,pol14}. The Stokes operators are nonlinear (quadratic) functions of the position and momentum operators of these two general modes. The reference mode in this Stokes-like measurement can simply be considered as another classical state (represented by a mixture of coherent states), which classically interferes with the signal mode. Identifying the state of the signal modes from Stokes operator measurements is generally challenging due to their non-linearity. The Stokes operator measurement approaches ideal homodyne detection only if a large-amplitude coherent state is injected to the probe mode \cite{bright1,bright2,bright3}. On the other hand, when the independent reference mode has vanishing mean values in any linear combination of position and momentum, it is not clear whether any or what kind of information about the covariance matrix can be estimated.

In this paper, we propose a method that efficiently estimates the covariance matrix of macroscopic quantum states from the Stokes operator measurements using {\em only classical, noisy and low intensity reference beams}. We identify a non-equilibrium character of such a reference as a necessary and sufficient condition for estimating the covariance matrix. 
The method can be directly applied to current experiments with bright squeezed vacuum \cite{Isk1,Isk2,Isk3}. 
This measurement strategy substantially relaxes the requirements for the reference addressing any SU(2) quantum measurement expressed in the Schwinger representation \cite{schwi}. 
For example, it can be extended to the spin squeezing of atomic ensembles \cite{spinsq1,spinsq2,spinsq3,spinsq4,spinsq5}.

The article is organized as follows. In Section II we describe a minimal measurement scheme needed for the estimation of the covariance matrix. In Section III we obtain an estimation procedure for the general case and investigate its properties. In Section IV we examine the possibilities with non-displaced references, while in Section V we summarize our results.


\section{Stokes-like measurements} Let us consider the simplest case: there is an independent signal ($S$) and a reference or ancilla ($R$) state. But in contrast to the standard homodyne measurement the signal is a macroscopic quantum state, that is, its energy is so high that it can be directly measured by a macroscopic detector (e.g., with a PIN-diode used in \cite{Isk1,Isk2,Isk3}). By approaching the macroscopic regime, the more and more powerful signals would require the use of extremely strong references, which can be challenging in practice. Therefore the focus of this work is to examine a scenario in which the reference has a magnitude of energy similar to or even lower than that of the signal. 
Also, instead of weak coherent states we examine arbitrary states as reference, that is, ones that can have zero or close to zero mean, or they may not be necessarily pure.

\begin{figure}[!t]
\begin{center}
  \includegraphics[width=6cm]{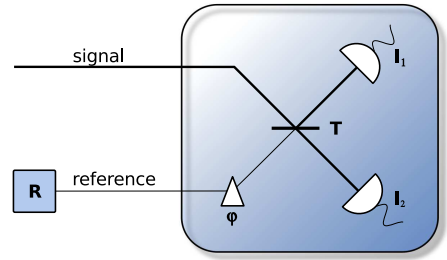}
 \caption{(Color online) Schematic figure of the Stokes-like measurement with an uncorrelated classical, noisy and low-intensity reference interfering with the macroscopically bright signal. T stands for the transmittance of a beam splitter, $\varphi$ is a phase shift between the signal and the reference, and $I_1$ and $I_2$ are photo-currents from standard intensity detectors.
  \label{Stokes}}
\end{center}
\end{figure}

The measurement setup for the homodyne measurement is generalized to a Stokes measurement (see Fig. \ref{Stokes}). The signal and the reference interfere at the beam splitter and the detectors give photocurrents proportional to the photon number averaged over coherence time and volume.

Without a beam splitter ($T=1$), we can access only the normally-ordered moments of photon numbers for the signal and the reference. Their sum and difference are proportional to the Stokes operators $(S_0,S_1)=a_S^\dagger a_S \pm  a_R^\dagger a_R=\frac{x_S^2+p_S^2-2}{4}\pm\frac{x_R^2+p_R^2-2}{4}$, where $a_i,a^{\dagger}_i$ $(i\in\{R,S\})$ are the annihilation and creation operators, and $x_i$, $p_i$ are the position and momentum operators of the signal and the reference modes. These satisfy commutation relations $[a_i,a_j^{\dagger}]=\delta_{ij}$ and $[x_i,p_j]=2i\delta_{ij}$.

We want to characterize the signal state, and for that purpose we estimate the mean vector $\mbox{m}_S$ and the covariance matrix $\mbox{V}_{S}$ of the signal,
\begin{equation}
\mbox{m}_S=\left(\<x_S\>,\<p_S\>\right),
\end{equation}
\vspace*{-0.7cm}
\begin{equation}
\mbox{V}_{S}=\left(\begin{array}{cc}
\<x_S^2\>-\<x_S\>^2 & \<x_Sp_S\>_s-\<x_S\>\<p_S\> \\
\<x_Sp_S\>_s-\<x_S\>\<p_S\> & \<p_S^2\>-\<p_S\>^2
\end{array}\right)
\end{equation}
where $\<x_Sp_S\>_s=(\<x_Sp_S\>+\<p_Sx_S\>)/2$.
Clearly, even all moments of photon numbers of the signal are insufficient to estimate these parameters, therefore we have to use an interference between the signal and the reference.

For a beam splitter with a transmittance of $T=0.5$ we implement the measurement of the Stokes operator
\begin{equation}
S_2=a_S^\dagger a_R + a_R^\dagger a_S=\frac12 (x_S \cdot x_R+ p_S \cdot p_R).
\end{equation}
Note that $\<S_2\>$, $\<S_2^2\>$ are functions of the first and second moments of the quadrature variables of the signal and the reference. Hence, by applying the inverse of these functions to the measurement outcomes it is possible to characterize the signal state. But even then we can not access to the off-diagonal elements of the covariance matrix. Thus, we apply an additional phase-shift ($\varphi$) to the reference to obtain the sufficiently generalized operators
\begin{equation}\label{S2phi}
S_2(\varphi)=\frac12 (x_S \cdot x_R^\varphi+ p_S \cdot p_R^\varphi),
\end{equation}
where $x_R^\varphi$ and $p_R^\varphi$ are the quadrature variables of the reference after the phase-shift. This operator is directly proportional to the difference of the two photo-currents ($I_1-I_2$) depicted in Fig. \ref{Stokes}. We should also remark that additional values of $T$ can be described as a combination of the cases with $T=0.5$ and $T=1$, therefore these would not improve the situation qualitatively.


\section{State estimation in general} 
\begin{figure}[!t]
\begin{center}
\includegraphics[width=6cm]{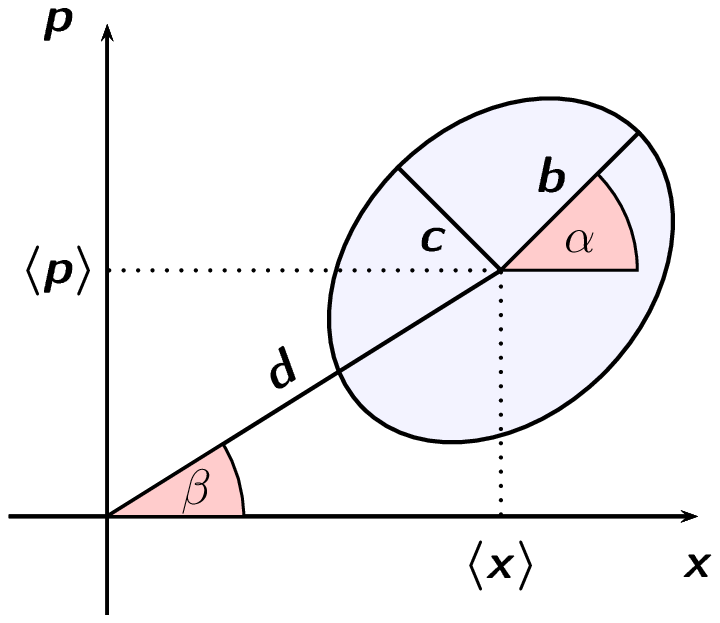}
 \caption{(Color online) The used parametrization of Gaussian states. One can get an arbitrary Gaussian state from a thermal state (with a proper standard deviation $r\ge 1$) using a squeezing and displacement. The shape of the Gaussian state is defined by squeezing: by using a quadrature squeezing in direction $\alpha$ with a magnitude of $q\ge 1$, we get the ellipse of the covariance matrix (note that $b=r\cdot q$ and $c=r/q$ are the square roots of the eigenvalues of the covariance matrix ($b\ge c$) and $\alpha$ defines its eigendirection). The displacement of the ellipse is given by its direction $\beta$ and its magnitude $d$  (note that this corresponds to the mean of the Gaussian state: $\<x\>=d\cos \beta,\<p\>=d\sin \beta)$. \label{ellipse}}
\end{center}
\end{figure}

Let us assume that we have two independent, general Gaussian states (Fig. \ref{ellipse}) for the signal and the reference (using subscripts $*_S$ and $*_R$, respectively).

If the reference is in a thermal state, then our estimation will be phase-insensitive. Thus, in order to introduce an asymmetry, we apply a squeezing (parametrized by $q,\alpha$) and/or a displacement (parametrized by $d,\beta$) to a thermal state (parametrized by $r$) to obtain classical states out of thermal equilibrium, as depicted in Fig. \ref{ellipse}.
For the sake of simplicity we investigate the case when the directions of squeezing and displacement in the reference coincide (if both are present), and we use this direction as a phase reference (that is, $\beta_R=\alpha_R=0$). In the further discussion we assume that we know the other parameters ($b_R,c_R,d_R$) of the reference, i.e., we already performed some a priori calibration. 


\subsection{General estimation method} 

For state reconstruction we will use only the first and second moments of $S_2(\varphi)$ from (\ref{S2phi}).
To obtain sufficient information, one should perform measurements in three different directions. For the sake of simplicity we will use unaltered ($\varphi=0$), orthogonal ($\varphi=\pi/2$) and diagonal ($\varphi=\pi/4$) references. 
Note that $S_2(0)$ coincides with Stokes operator $S_2$, $S_2(\pi/2)$ coincides with Stokes operator $S_3$, but $S_2(\pi/4)$ (that estimates $\Cov(x_S,p_S)$) is not a standard Stokes operator.

The estimation of the displacement of the signal is straightforward. If we leave the reference unaltered ($\varphi=0$), we get the mean of the $x$-quadrature directly:
\begin{equation}
\<S_2(0)\>=\<x_S\> \cdot d_R/2+\<p_S\> \cdot 0 \Rightarrow \<x_S\>=\frac{2\<S_2(0)\>}{d_R}.
\end{equation}
Similarly, if we apply a phase shift of $\varphi=\pi/2$ to the reference, we obtain the mean of the $p$-quadrature:
\begin{equation}
\<S_2(\pi/2)\>=\<p_S\> \cdot d_R/2 \Rightarrow \<p_S\>=\frac{2\<S_2(\pi/2)\>}{d_R}.
\end{equation}
It is easy to see that the necessary and sufficient criteria for the feasibility of these estimates is $d_R\ne 0$. Also note that for other directions of the reference ($\varphi\ne 0, \varphi\ne \pi/2$), the value of $\<S_2\>$ can be calculated as a linear combination of the above equations.

The estimation of the variances is a little more complicated. We have two linear equations for the second moments:
\begin{equation}
\<S_2^2(0)\>=\<x_S^2\>\cdot (b_R^2+d_R^2)/4+\<p_S^2\>\cdot c_R^2/4+\frac12
\end{equation}
and
\begin{equation}
\<S_2^2(\pi/2)\>=\<x_S^2\> \cdot c_R^2/4+\<p_S^2\> \cdot (b_R^2+d_R^2)/4+\frac12.
\end{equation}
We can solve these equations uniquely and obtain the values of $\<x_S^2\>$ and $\<p_S^2\>$.
From these second moments one can easily calculate the variances: $\Var(x_S)=\<x_S^2\>-\<x_S\>^2$ and $\Var(p_S)=\<p_S^2\>-\<p_S\>^2$.

Finally, the covariance can be calculated from the second moment using a diagonal measurement ($\varphi=\pi/4$). We have
\begin{equation}
\begin{split}
\<S_2^2(\pi/4)\>&=\frac12+\<x_S^2\> \cdot (d_R^2+b_R^2+c_R^2)/8+\\
&+\<p_S^2\> \cdot (d_R^2+b_R^2+c_R^2)/8+\\
&+ \<x_S p_S\>_s \cdot (d_R^2+b_R^2-c_R^2)/4.
\end{split}
\end{equation}
Since we already know $\<x_S^2\>$ and $\<p_S^2\>$ we can calculate $\<x_S p_S\>_s$ easily, and then we get the covariance by $\Cov(x_S,p_S)=\<x_S p_S\>_s- \<x_S\> \<p_S\>$. 

Note that the above method in principle coincides with one of the standard estimation methods using homodyne measurement. In that case we have $b_R=c_R=1$ and $d_R\gg1$, so all terms which do not contain $d_R$ will be negligible. Thus, $\<S_2^2(\varphi)\>$ will define approximately the second moment of the signal in the direction of the reference ($\varphi$). 


\subsection{Alternative parametrization} To gain a better understanding of the estimation using the structure of Stokes measurements, we can use the eigendecomposition of the covariance matrix ($b$, $c$ and $\alpha$) instead of its elements ($\Var(x), \Var(p)$ and $\Cov(x,p)$). The connection between the two parametrizations can be described by the following equations:
\begin{eqnarray}
\Var(x)=b^2 \cos^2(\alpha)+c^2 \sin^2(\alpha)\\
\Var(p)=c^2 \cos^2(\alpha)+b^2 \cos^2(\alpha)\\
\Cov(x,p)=(b^2-c^2) \cos(\alpha) \sin(\alpha)
\end{eqnarray}

In this case the second moment of $S_2(\varphi)$ will have the following form:

\begin{equation}\label{S22phi}
\begin{split}
\<S_2^2(\varphi)\>&=\frac18 (d_R^2+b_R^2+c_R^2)(d_S^2+b_S^2+c_S^2)+\\ 
 &+\frac18 (d_R^2+b_R^2-c_R^2)(b_S^2-c_S^2) \cos(2\alpha_S-2\varphi)+\\
 &+\frac18 (d_R^2+b_R^2-c_R^2) d_S^2\cos(2\beta_S-2\varphi).
\end{split}
\end{equation}

That is, it consists of three parts: 
\begin{enumerate}
\item the total energy of the signal, 
\item the asymmetry induced by the shape of the signal,
\item the asymmetry induced by the displacement of the signal.
\end{enumerate}

From the first moments we can calculate the parameters $d_S$ and $\beta_S$. 
Using these the last term becomes known and we have
\begin{equation}\label{sqsq}
\<S_2^2(\varphi)\>=u(b_S^2+c_S^2)+v(b_S^2-c_S^2)\cos(2\alpha_S-2\varphi)+w,
\end{equation}
where $u=\frac18 (d_R^2+b_R^2+c_R^2)$, $v=\frac18 (d_R^2+b_R^2-c_R^2)$ and $w$ is a constant depending on the reference, $d_S$ and $\beta_S$.

The total energy is directly accessible from $\<S_0\>$, or equivalently by using two orthogonal references:
\begin{equation}\label{sqsq_x}
\<S_2^2(0)\>=u(b_S^2+c_S^2)+v (b_S^2-c_S^2) \cos(2\alpha_S)+w,
\end{equation}
\begin{equation}\label{sqsq_p}
\<S_2^2(\pi/2)\>=u(b_S^2+c_S^2)-v (b_S^2-c_S^2) \cos(2\alpha_S)+w.
\end{equation}

The estimation of the asymmetry of the ellipse ($b_S^2-c_S^2$) and its direction ($\cos(2\alpha_S)$) only appear as a product with each other. So to access these parameters separately, we need an additional equation. For simplicity we will use a diagonal reference:
\begin{equation}\label{sqsq_diag}
\<S_2^2(\pi/4)\>=u(b_S^2+c_S^2)+v (b_S^2-c_S^2) \sin(2\alpha_S)+w,
\end{equation}
but in principle any other angle ($0<\varphi<\pi/2$) is equally adequate.

From equations (\ref{sqsq_x})-(\ref{sqsq_diag}) one can already calculate the parameters of the signal. The constant $u$ is always greater than zero, so the necessary and sufficient criterion for the feasibility of these estimates is $v\ne 0$. This holds if the reference is not thermal. Note that $\<S_2^2(\varphi)\>$ in Eq.(\ref{sqsq}) is a linear transformation of a cosine function, so it has three parameters. Equations (\ref{sqsq_x})-(\ref{sqsq_diag}) uniquely characterize this function, that is, including a fourth angle for reference would not give any additional information. One of the equations from (\ref{sqsq_x}) and (\ref{sqsq_p}) is interchangeable with
\begin{equation}
\<S_0\>=\frac{b_S^2+c_S^2+d_S^2}{4}+\frac{b_R^2+c_R^2+d_R^2}{4}-1.
\end{equation}
Only the moments $\<S_0\>, \<S_2(\varphi)\>, \<S_2^2(\varphi)\>$ can be described as the first and second moments of the signal, and based on the above discussion we can see that there are at most five independent equations amongst them for different values of $\varphi$. A general Gaussian state has also five parameters, so the proposed state estimation method is tight in this sense. 


\subsection{Properties of the estimation}

\begin{figure}[!t]
\includegraphics[width=0.48\columnwidth]{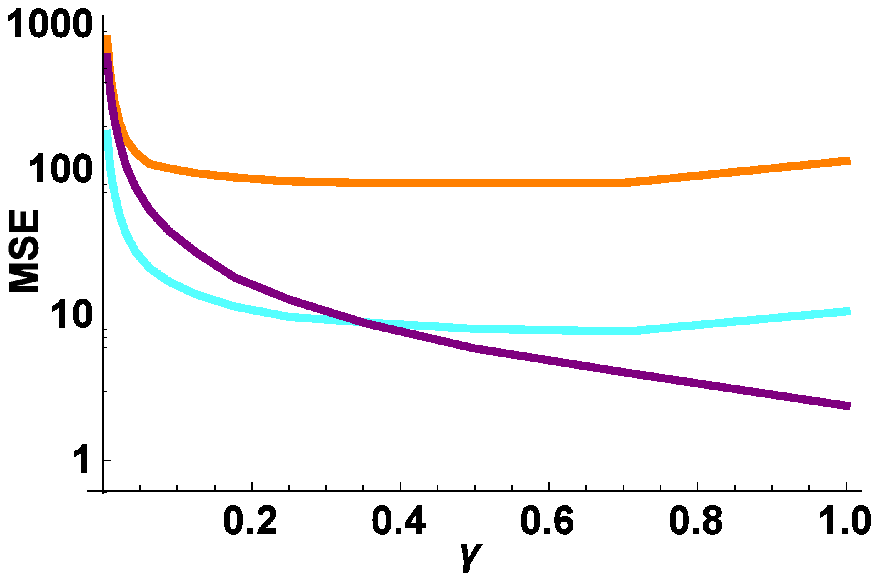}
\includegraphics[width=0.48\columnwidth]{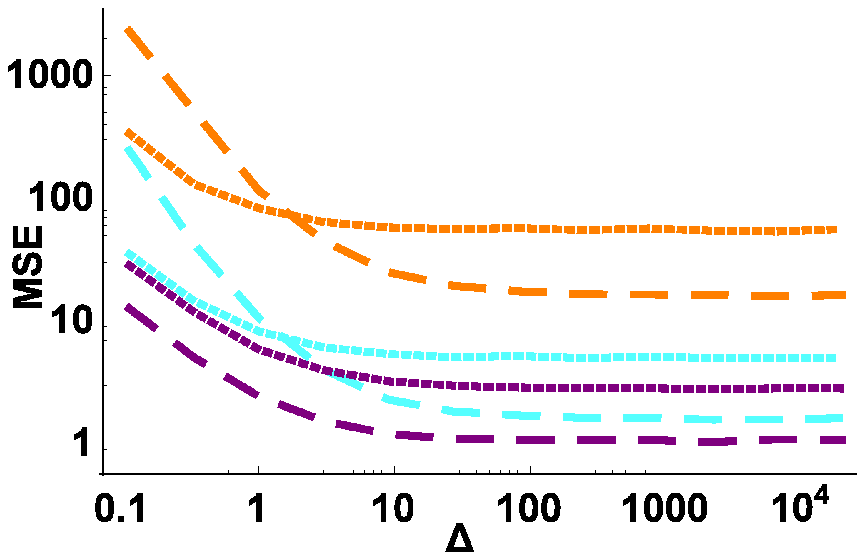}
\caption{\label{general}(Color online) MSE of the state estimation of a general signal as a function of the displacement ratio (left figure) and NER (right figure) of the reference. Cyan (light) lines correspond to the estimation of parameter $b$, orange (medium) to parameter $c$ and purple (dark) to parameter $d$. In the left figure we used $\Delta=1$. In the right figure we used both displaced thermal reference (dashed lines, $\gamma=1$) and displaced$+$squeezed thermal reference (dotted lines, $\gamma=0.5$). The MSE is calculated using $N=10^5$ Gaussian states and signal parameters: $b_S=237, c_S=86,\alpha_S=0.7, d_S=158, \beta_S=0.2$. Note that the figures look qualitatively the same for different parameters.}
\end{figure}

To quantify the quality of the estimation, we use the mean squared error (MSE):
\begin{equation}
\mbox{MSE}_N(\hat\theta)=\<(\hat\theta-\theta)^2\>,
\end{equation}
where $\theta$ is an estimated parameter (it can be: $d_S$, $b_S$, $c_S$, $\alpha_S$, $\beta_S$), $\hat{\theta}$ is the estimator of this parameter and $N$ is the number of used signal states. If the estimation is feasible, we have an asymptotic behavior of $\mbox{MSE}_N(\hat\theta)\sim 1/N$, multiplied by a coefficient which depends on the parameters of the signal and the reference. The signal is inaccessible, so in the following we will investigate how the MSE depends on the parameters of the reference. Since our scheme has many parameters which are the variables in the formula of the MSE, the analytic formula is quite lengthy and not very informative. So instead of using that, we simply plot the empirical MSE from $10^4$ numerical simulations of the estimation process. Let us also note that our estimator can provide a non-physical state, but for a large number of measurement the probability of this event decreases exponentially, so using restrictions on the estimator would have a negligible effect on MSE.

To characterize non-equilibrium features of the reference, we introduce the following \textit{non-equilibrium energy ratio} (NER):
\begin{equation}
\Delta=\frac{n_{\text{total}}-n_{\text{thermal}}}{n_{\text{thermal}}}.
\end{equation}
This quantity describes how the energy added to the symmetrical reference (to move it away from equilibrium) relates to the energy of the original, symmetrical state. For a reference that is only displaced ($q=1$) we have $\Delta_{\text{disp}}=\frac12 \frac{d^2}{r^2}$, while for a reference that is only squeezed ($d=0$) we have $\Delta_{\text{sq}}=\frac12 (q-1/q)^2$. Note that for a squeezed and displaced state we have $\Delta=\Delta_{\text{disp}}+\Delta_{\text{sq}}$, that is, we can define a \textit{displacement ratio}: $\gamma=\Delta_{\text{disp}}/\Delta$ ($0\le\gamma\le1$), which describes how much of the non-equilibrium characteristic comes from the displacement. This way the triplet of $(r_R,\Delta,\gamma)$ parametrizes the reference states uniquely. 

Our first observation is that the MSE of the estimators does not depend on $r_R$. The dependence on the displacement ratio can be seen in the left part of Fig.~\ref{general}. If we only apply squeezing to the reference \textit{we cannot estimate the signal}, the MSE diverges as $\gamma\rightarrow 0$. However, if there is also an arbitrary small displacement present in the reference ($\gamma>0$), then from the asymptotic behavior of the MSE it follows that the error converges to zero (i.e., our estimations converge to the real parameters) as the number of used Gaussian states ($N$) converges to infinity. The dependence on $\gamma$ is not trivial, usually a purely displaced reference ($\gamma$=1) is the best scenario, but adding some squeezing to the reference can improve the estimation.

This effect is also visible in the right subfigure of Fig.~\ref{general}, where at estimating parameter $b$ and $c$ the squeezed+displaced (dotted lines) reference outperforms the displaced reference (dashed lines) for small NER (i.e., when the reference is close to a thermal state). More importantly, we can also conclude that the estimation of the signal is possible even for an arbitrary small deviation of the Gaussian reference from thermal equilibrium. If NER ($\Delta$) increases the estimation always becomes more efficient. The displaced reference always saturates at the lowest level, providing the best estimation for strongly asymmetrical references (this explains the popularity of using coherent states as a strong local oscillator). However, this saturation happens at a relatively low level of asymmetry ($\Delta \sim 10-100$). That is, from the estimation point of view, there is \textit{no principal need to use the standard strong coherent reference} (e.g., $\Delta\sim 10^5$). The macroscopic signal in itself provides enough energy for the detectors, so even with a weak reference with a little asymmetry a precise estimation is achievable. Note, that we plot only $b_S$, $c_S$ and $d_S$ to avoid too many lines in the graphs, the estimation of $\alpha_S$ shows strong similarities with the estimation of $b_S$ and $c_S$, and the estimation of $\beta_S$ is similar to $d_S$.

\subsection{Robustness of the estimation}

From the previous subsection we know that the obtained estimators are always asymptotically unbiased if the reference is precisely known. 
In the following we will investigate how this changes when the reference is not an ideal, known Gaussian state. 

The first possibility is that the reference is not fixed, but rather fluctuates around a value. 
In this case the reference will be the superposition of the possible states, which, in general, will not produce a Gaussian state. 
This is, however, not generally a problem because the used estimators do not rely on the Gaussianity of the states. 
We use the first and second moments of the reference, and if we have precise estimates of those our method always produces asymptotically unbiased estimates, even in this non-Gaussian scenario.

\begin{figure}[!t]
\begin{center}
\includegraphics[width=0.48\columnwidth]{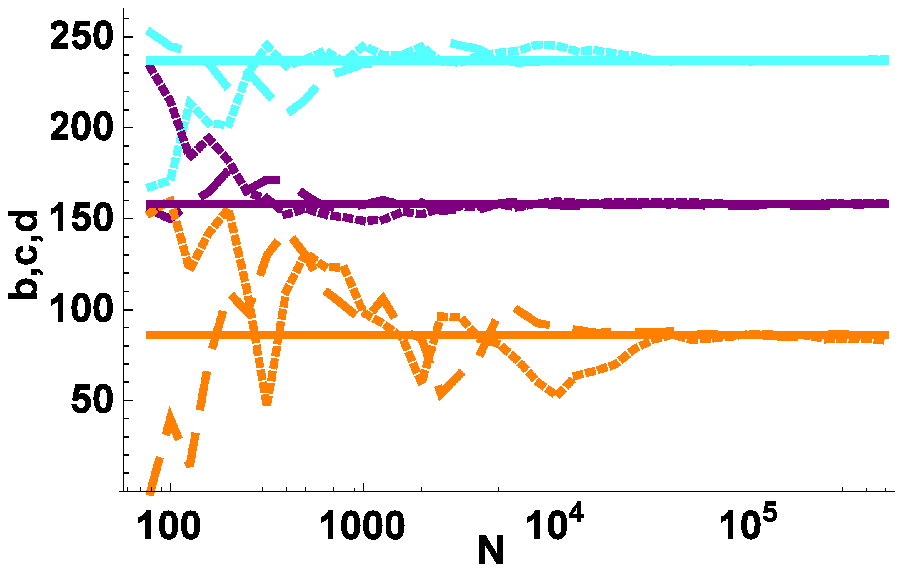}
\includegraphics[width=0.48\columnwidth]{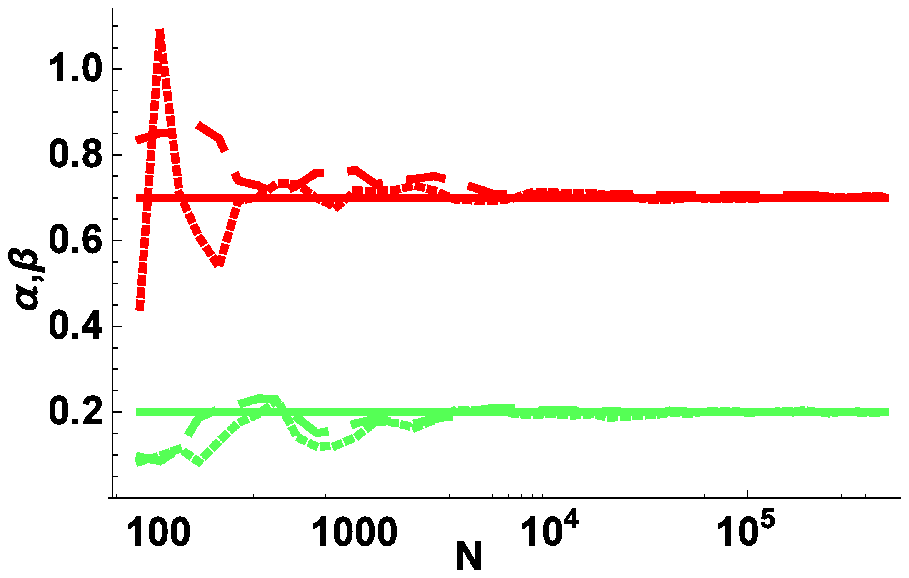}
\caption{(Color online) Estimates of signal parameters as a function of the number of measurements. In the left figure the cyan (light) lines correspond to the estimates of parameter $b$, orange (medium) to parameter $c$ and purple (dark) to parameter $d$. In the right figure the red (dark) lines correspond to the estimates of parameter $\alpha$ and green (light) to parameter $\beta$. We used both displaced thermal reference (dashed lines, $\gamma=1$) and displaced$+$squeezed thermal reference (dotted lines, $\gamma=0.5$) and compared the estimates with the real values of the parameter (solid lines). The estimates are calculated using $\Delta=10$ and signal parameters: $b_S=237, c_S=86,\alpha_S=0.7, d_S=158, \beta_S=0.2$.\label{number}}
\end{center}
\end{figure}

That is, we only get an imprecise estimation of the signal state if the moments of the reference are imprecise. A bias in the reference direction (i.e., $\alpha_R=\beta_R=\varepsilon\ne 0$) simply results in the same bias in the estimation of the signal direction, i.e., in $\alpha_S$ and $\beta_S$, and would not influence the non-angle parameters ($b_S,c_S,d_S$). The opposite statement is also true, that is, the bias in non-angle parameters will result in an error only in the non-angle parameters of the signal states. However, in contrast to the angle-parameters, this relation is not straightforward owing to the non-linearity of the non-angle parameter estimators. 

The most common example for this bias appears when we have a finite number of measurements. In this case even if the expected values of the moments are known, for an actual realization the moments will be different. This will result in a biased estimation of the signal state, but we can see from Fig. \ref{number} that for about $10^4-10^5$ measurements the difference vanishes. By this natural example, we clearly demonstrate that the problem of reference parameter accuracy becomes negligible even for a relatively low number of measurements.


\section{State estimation with squeezed thermal reference} 

We have seen that state tomography is feasible in general only in the case of a displaced reference ($d_R>0$), so in the following we will investigate the case of non-displaced reference ($d_R=0$).

\subsection{Possibilities in the general case} A non-displaced squeezed reference has zero mean in quadratures (i.e., $\<x_R^\varphi\>=\<p_R^\varphi\>=0$), therefore the mean of operator $S_2$ will be zero:
\begin{equation}\label{zero}
\<S_2(\varphi)\>=1/2 \<x_S\>\<x_R^\varphi\>+1/2\<p_S\>\<p_R^\varphi\>=0.
\end{equation}

\begin{figure}[!ht]
\begin{center}
  \includegraphics[width=8.5cm]{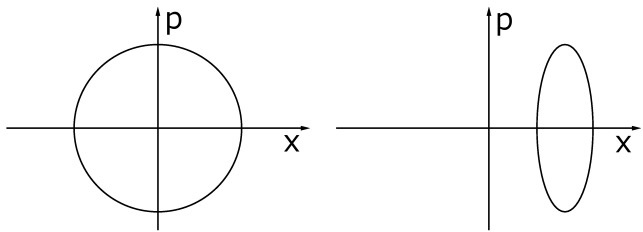}
 \caption{Different variants of the signal state with the same second moments, therefore they are indistinguishable by using an only squeezed reference. \label{mean}}
\end{center}
\end{figure}

So one can use only $\<S_0\>$ and $\<S_2^2(\varphi)\>$ for state estimation, which results in three independent equations. That is, with a squeezed reference the estimation of the signal is not possible in general. Since $\<S_0\>$ and $\<S_2^2(\varphi)\>$ are functions of the second moments one cannot distinguish for example a thermal state from a displaced squeezed state with the same second moments (see Fig.\ \ref{mean}). 

By assuming that the state is Gaussian we can use also $\<S_0^2\>$ for estimation and with its help we can discriminate for example the two cases shown in Fig.\ \ref{mean}. Nevertheless, even in that case the estimation is not unique. For example, let us consider a displaced thermal state as a signal: $b_S=c_S=2$, $\<x_S\>=3$ and $\<p_S\>=0$. The values of $\<S_2^2(\varphi)\>$ and $\<S_0^2\>$ will be the same for: $b_S\approx 1.5118$, $c_S\approx 2.3905$, $\<x_S\>\approx 2.9451$, $\<p_S\>\approx 0.5714$ and $\alpha_S\approx 4.0228$. That is, even using $\<S_0^2\>$ there are states which are indistinguishable. However, one can see that the values of $b_S^2+c_S^2$ and $d_S^2=\<x_S\>^2+\<p_S\>^2$ are the same for all the possibilities. By these measurements we can at least discriminate the energy coming from the displacement and from the internal variance of the signal. But we cannot determine the internal asymmetry (ratio of $b_S$ and $c_S$) of the source (actually it could be anything within the given energy constraint). 


\subsection{State estimation in special cases} 
We showed that the general state estimation is not feasible for a non-displaced reference, however, state tomography is viable in some special cases.

\begin{figure}[!t]
\includegraphics[width=0.48\columnwidth]{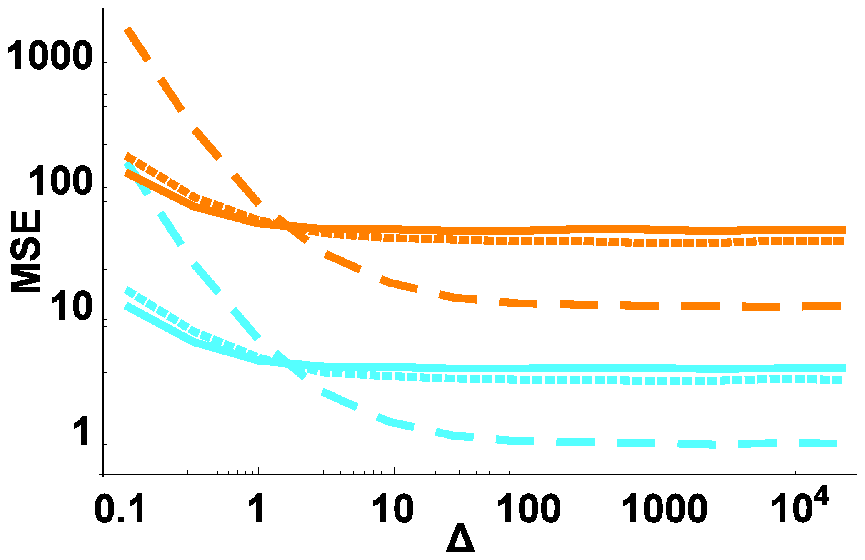}
\includegraphics[width=0.48\columnwidth]{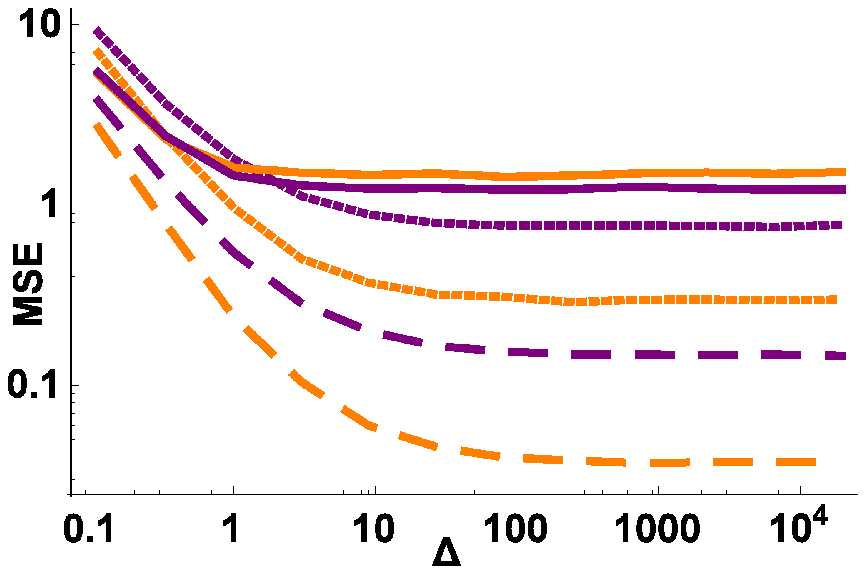}
\caption{\label{special} (Color online) MSE of the state estimation of a non-displaced squeezed signal (left figure) and a symmetrical displaced signal (right figure) as a function of the NER of the reference. Cyan (light) lines correspond to the estimation of parameter $b$, orange (medium) to parameter $c$ and purple (dark) to parameter $d$. We used displaced thermal reference (dashed lines, $\gamma=1$), displaced$+$squeezed thermal reference (dotted lines, $\gamma=0.5$) and classical squeezed thermal reference (solid lines, $\gamma=0$). The MSE is calculated using $N=10^5$ Gaussian states. For squeezed state we have $b_S=237, c_S=86, \alpha_S=0.7, d_S=0$, for displaced state we have $b_S=c_S=86, d_S=158, \beta_S=0.2$.}
\end{figure}

If we know that the signal is only squeezed (i.e., $d_S=0$), even using an only squeezed reference does not present any difficulties, since we do not need to estimate the first moments. And as it is mentioned in the previous section, the estimation of the second moments is feasible if the reference is not thermal. There is not much difference compared to a general signal (left subfigure of Fig. \ref{special}), because knowing the displacement is not a substantial advantage (since estimating the second moments is generally more difficult than estimating the first moments). But since we do not need to estimate the displacement, we can perform an estimation with a purely squeezed ($\gamma=0$) reference, too. This will be the best option in the case of a small asymmetry of the reference, while the worst in the case of a large asymmetry.

If we know that the signal is only displaced (i.e., $b_S=c_S=r_S$), Eq. (\ref{S22phi}) reduces to
\begin{equation}\label{S22_sqsqsq}
\begin{split}
\<S_2^2(\varphi)\>&=\frac18 (b_R^2+c_R^2)(d_S^2+2 r_S^2)+\\ 
  &+\frac18 (b_R^2-c_R^2) d_S^2\cos(2\beta_S-2\varphi).
\end{split}
\end{equation}
which has the same structure as Eq. (\ref{sqsq}) with $b_S^2+c_S^2\leftrightarrow d_S^2+2 r_S^2$, $b_S^2-c_S^2\leftrightarrow d_S^2$ and $\alpha_S\leftrightarrow \beta_S$ therefore it can be solved similarly. Once again, the estimation is feasible if $b_R\ne c_R$. Knowing that the covariance matrix is a multiple of the identity largely improves its estimation (right subfigure of Fig. \ref{special}). Interestingly, in this case we can even estimate the displacement with a squeezed reference. The efficiency of the estimators is worse compared to the displaced reference. The reason is simple: for the displaced reference one can obtain the displacement of the signal directly from the first moments, while for the squeezed reference we can only obtain it indirectly from the second moments.


If the reference is symmetrical ($b_R=c_R=r_R$ and $d_R=0$) all estimates diverge, that is, with a thermal reference we cannot estimate the parameters even in these special cases. In this case $\<S_2^2(\varphi)\>$ will be equivalent for any $\varphi$ with $\<S_0\>$, so we can only access the energy of the signal: 
\begin{equation}\label{th}
\<S_0\>=\frac{b_S^2+c_S^2+d_S^2}{4}+\frac{r_R^2}{2}-1.
\end{equation}

However, if we know that the signal is Gaussian, we can use higher moments as well. The perfect candidate for that is $S_0^2$, which is still the second moment of the measurement results, hence it converges relatively quickly to its mean. And its mean consists of fourth moments of quadratures (beside the second moments), which in the Gaussian case can be described as functions of second moments. This gives us an additional equation, and with that we can solve some special cases of estimation problems. 

If the signal is a squeezed, non-displaced Gaussian state, then 
\begin{equation}
\begin{split}
\<S_0^2\>&=\frac{3 b_S^4+2b_S^2 c_S^2+3c_S^4}{16}+\\
&+(b_S^2+c_S^2) (r_R^2-2)/4+f(r_R).
\end{split}
\end{equation}
Combining this with Eq. (\ref{th}) (using also $d_S=0$) one can calculate the values of $b_S$ and $c_S$.

We have for a displaced, symmetrical Gaussian signal
\begin{equation}
\begin{split}
\<S_0^2\>&=\frac{d_S^4+8d_S^2 r_S^2+ 8r_S^4}{16}+\\
&+(d_S^2+2r_S^2) (r_R^2-2)/4+f(r_R).
\end{split}
\end{equation}
and once again combining with Eq. (\ref{th}) (using $b_S=c_S=r_S$), the values of $r_S$ and $d_S$ can be calculated . 

Note that we determined in the above cases the magnitude of squeezing and the magnitude of displacement, but not their direction. That is not surprising since the reference is symmetrical, so the measurement outcomes show the same statistics for any angle. Note that even if the states are not Gaussian, by using multi-copy interference Gaussification process, they can get arbitrary close to Gaussian \cite{Gauss}.


\section{Conclusion and Discussion} We investigated the case of measuring a macroscopic signal. In this case the standard homodyne measurement would need an extremely strong local oscillator, but in principle a classical, noisy and low intensity reference is sufficient. It was not known what are the minimal conditions for a successful estimation of the covariance matrix in this situation. Our work shows that if there is at least a small displacement from the thermal equilibrium in the reference, a full tomography of the signal is always possible. The quality of the estimation highly depends on the properties of the reference state, mainly on its non-equilibrium energy ratio (NER): one can obtain a better estimation if the reference is further away from thermal equilibrium. The efficiency saturates at a low level of asymmetry, so there is no fundamental need to use strong local oscillators, even if such are available. If the reference is really close to a thermal state, then we can still reasonably estimate the signal, we should only use more Gaussian states for estimation. There are also differences in behavior for different types of non-equilibrium nature of the reference. The displaced reference in most cases gives the best or close to best performance, but some additional squeezing of the reference can even improve that. 

The current paper investigates only the existence and main characteristics of such a scheme, so a possible future direction is to improve these results with more elaborate techniques. The proposed detection technique allows to detect macroscopic quantum states of light without a coherent local oscillator, which can lead to interesting applications in quantum communication and metrology.  One such direct application is continuous-variable quantum key distribution with macroscopic squeezed states of radiation \cite{Isk1,Isk2,Isk3,Isk4}, where the estimation of the covariance matrix is necessary to obtain a secure key rate \cite{Mad12}. Another possible application is to estimate the spin squeezing of atomic ensembles \cite{spinsq3,spinsq4,spinsq5} by using similar techniques. 

\section*{Acknowledgment} V.C.U. and L.R. acknowledge the Project No. 13-27533J of the Czech Science Foundation. The research leading to these results has received funding from the EU FP7 under Grant Agreement No. 308803 (Project BRISQ2), co-financed by M\v SMT \v CR (7E13032). L.R. acknowledges support by the Development Project of Faculty of Science, Palacky University.


\begin{thebibliography}{0}

\bibitem{Peres}
A. Peres, Quantum Theory: Concepts and Methods, Kluwer Academic Publishers,
(1993).

\bibitem{Paris}
M. Paris and J. \v Reh\'a\v cek, eds. Quantum State Estimation, Springer, Vol. 649., (2004).

\bibitem{Sakurai}
J.J. Sakurai, Modern Quantum Mechanics, Addison-Wesley (1991).

\bibitem{covmat}
Ch. Weedbrook, S. Pirandola, R. Garcia-Patron, N.J. Cerf, T.C. Ralph, J.H. Shapiro, and S. Lloyd, Rev. Mod. Phys. 84, 621 (2012).

\bibitem{squeezing1}
D.R. Robinson, Commun. Math. Phys. 1, 159 (1965).

\bibitem{squeezing2}
D. Stoler, Phys. Rev. D1, 3217 (1970); D4, 1925 (1971).

\bibitem{noncl}
L. Mandel and E. Wolf, Optical Coherence and Quantum Optics, Cambridge University Press, (1995).


\bibitem{Nav}
M. Navascues, F. Grosshans and A. Acin, Phys. Rev. Lett. 97, 190502 (2006).

\bibitem{Gar}
R. Garcia-Patron and N. J. Cerf, Phys. Rev. Lett. 97, 190503 (2006).

\bibitem{Mad12}
L. S. Madsen, V. C. Usenko, M. Lassen, R. Filip and U. L. Andersen, Nat. Com. 3, 1083 (2012). 

\bibitem{Rup14}
V. C. Usenko, L. Ruppert and R Filip, Phys. Rev. A 90(6), 062326 (2014).

\bibitem{Morgan1}
F. Wolfgramm, A. Cere, F.A. Beduini, A. Predojevic, M. Koschorreck, M.W. Mitchell, Phys. Rev. Lett. 105, 053601 (2010). 

\bibitem{Luis1}
\' A. Rivas and A. Luis, Phys. Rev. Lett. 105, 010403 (2010)

\bibitem{Morgan2}
R. J. Sewell, M. Koschorreck, M. Napolitano, B. Dubost, N. Behbood, and M. W. Mitchell, 
Phys. Rev. Lett. 109, 253605 (2012).

\bibitem{Genovese}
E. D. Lopaeva, I. Ruo Berchera, I. P. Degiovanni, S. Olivares, G. Brida, and M. Genovese, 
Phys. Rev. Lett. 110, 153603 (2013).

\bibitem{Brask1}
J.B. Brask, R. Chaves, and J. Kolodynski,
Phys. Rev. X 5, 031010 (2015).

\bibitem{Yuen}
H. P. Yuen and V. W. S. Chan, Opt. Lett. 8, 177, 345(E) (1983).

\bibitem{Schu}
B. L. Schumaker, Opt. Lett. 9, 189 (1984).

\bibitem{Isk1}
T. Sh. Iskhakov, M. V. Chekhova, G. O. Rytikov, and G. Leuchs, Phys. Rev. Lett. 106, 113602 (2011).

\bibitem{Isk2}
T.Sh. Iskhakov, I.N. Agafonov, M.V. Chekhova, and G. Leuchs, Phys. Rev. Lett. 109, 150502 (2012).

\bibitem{Isk3}
B. Kanseri, T. Iskhakov, I. Agafonov, M. Chekhova, and G. Leuchs,
Phys. Rev. A 85, 022126 (2012) 

\bibitem{Isk4}
B. Kanseri, T. Iskhakov, G. Rytikov, M. Chekhova, and G. Leuchs,
Phys. Rev. A 87, 032110 (2013).

\bibitem{Use}
V. C.  Usenko, L. Ruppert, R. Filip, Optics Express 23 (24), 31534 (2015).

\bibitem{Stok1}
G. G. Stokes, Trans. Cambridge Philos. Soc. 9, 399 (1852) and
Mathematical and Physical Papers (Cambridge, 1901), Vol. 3.

\bibitem{Stok2}
J. Schwinger, On angular momentum Quantum Theory of Angular Momentum, eds. L.C. Biedenharn and H. Dam, New York: Academic (1965).

\bibitem{Stok3}
A. Luis and L.L. S\' anchez-Soto, Progress in Optics 41, 421 (2000).

\bibitem{Vogel}
W. Vogel, Physical Review A 51(5), 4160 (1995).

\bibitem{Vogel2}
S. Wallentowitz and W. Vogel, Phys. Rev. A 53, 4528 (1996).

\bibitem{Opat}
T. Opatrn\'y and D.-G. Welsch, Phys. Rev. A 55, 1462 (1996).

\bibitem{Sanchez}
A. Cives-Esclop, A. Luis and L.L. S\' anchez-Soto, Optics Communications 175, 153–161 (2000).

\bibitem{Jiang}
L. A. Jiang and J.X. Luu, Appl. Opt. 47, 1486-1503 (2008). 

\bibitem{pol1}
D. M. Klyshko, Zh. Eksp. Teor. Fiz. 111, 1955 (1997). 

\bibitem{pol2}
N. Korolkova, G. Leuchs, R. Loudon, T.C. Ralph, and Ch. Silberhorn
Phys. Rev. A 65, 052306 (2002). 

\bibitem{pol3}
W.P. Bowen, R. Schnabel, H.-A. Bachor, and P.K. Lam
Phys. Rev. Lett. 88, 093601 (2002). 

\bibitem{pol4}
A. Luis, Phys. Rev. A 66, 013806 (2002). 

\bibitem{pol5}
A. Luis and N. Korolkova,
Phys. Rev. A 74, 043817 (2006).

\bibitem{pol6}
A.B. Klimov, J. Delgado, L.L. S\'anchez-Soto, Opt. Commun. 258, 210 (2006). 

\bibitem{Rivas2}
\' A. Rivas and A. Luis Phys. Rev. A 78, 043814 (2008)

\bibitem{pol7}
G. Bj\"ork, J. S\"oderholm, L. L. S\'anchez-Soto, A. B. Klimov, I. Ghiu, P. Marian, T. A. Marian,  Opt. Commun. 283, 4440 (2010).

\bibitem{pol8}
J S\"oderholm, G Bj\"ork, A.B. Klimov, L.L. S\'anchez-Soto and G. Leuchs, N.J. Phys. 14, 115014 (2012). 

\bibitem{pol9}
G. Bj\"ork, J. S\"oderholm, Y.-S. Kim, Y.-S. Ra, H.-T. Lim, C. Kothe, Y.-H. Kim, L. L. S\'anchez-Soto, and A. B. Klimov
Phys. Rev. A 85, 053835 (2012). 

\bibitem{pol10}
M. Stobinska, F. T\" oppel, P. Sekatski, and M.V. Chekhova
Phys. Rev. A 86, 022323 (2012). 

\bibitem{pol11}
P. de la Hoz, A. B. Klimov, G. Bj\"ork, Y.-H. Kim, C. M\"uller, Ch. Marquardt, G. Leuchs, and L. L. S\'anchez-Soto
Phys. Rev. A 88, 063803 (2013). 

\bibitem{pol12}
Ch. Kothe, L.S. Madsen, U.L. Andersen, and G. Bj\" ork,
Phys. Rev. A 87, 043814 (2013).

\bibitem{pol13}
G. Bj\" ork, H. de Guise, A. B. Klimov, P. de la Hoz, and L. L. S\' anchez-Soto, Phys. Rev. A 90, 013830 (2014). 

\bibitem{pol14}
G. Bj\" ork, A. B. Klimov, P. de la Hoz, M. Grassl, G. Leuchs, and L. L. S\' anchez-Soto,
Phys. Rev. A 92, 031801(R) (2015). 

\bibitem{bright1}
Ch. Marquardt, J. Heersink, R. Dong, M.V. Chekhova, A.B. Klimov, L.L. S\' anchez-Soto, U.L. Andersen, G. Leuchs, Phys. Rev. Lett. 99, 220401 (2007). 

\bibitem{bright2}
A. B. Klimov, G. Bj\" ork, J. S\" oderholm, L. S. Madsen, M. Lassen, U. L. Andersen, J. Heersink, R. Dong, Ch. Marquardt, G. Leuchs, and L. L. S\'anchez-Soto
Phys. Rev. Lett. 105, 153602 (2010).

\bibitem{bright3}
C.R. M\" uller, B. Stoklasa, C. Peuntinger, C. Gabriel, J. \v Reh\' a\v cek, Z. Hradil, A.B. Klimov, G. Leuchs, Ch. Marquardt, and L.L. S\' anchez-Soto, New J. Phys. 14, 085002 (2012). 

\bibitem{schwi}
J. Schwinger, Proc. Nat. Acad. Sci. U.S.A 46, 570 (1960).

\bibitem{spinsq1}
K. Hammerer, A.S. Sorensen, and E.S. Polzik, Review of Modern Physics, 82, 1041 (2010).

\bibitem{spinsq2}
J. Ma, X. Wang, C.P. Sun, F. Nori, Physics Reports 11, 509 (2010).

\bibitem{spinsq3}
F.A. Beduini and M.W. Mitchell, Phys. Rev. Lett. 111, 143601 (2013).

\bibitem{spinsq4}
N. Behbood, F. Martin Ciurana, G. Colangelo, M. Napolitano, G. T\'oth, R. J. Sewell, and M. W. Mitchell, Phys. Rev. Lett. 113, 093601 (2014).

\bibitem{spinsq5}
F.A. Beduini, J.A. Zieli\'nska, V.G. Lucivero, Y.A. de Icaza Astiz, and M.W. Mitchell, Phys. Rev. Lett. 114, 120402 (2015).

\bibitem{Gauss}
M.M. Wolf, G. Giedke, and J.I. Cirac, Phys. Rev. Lett. 96, 080502 (2006).

\end{thebibliography}
\end{document}